# Bayesian approach to superstatistics

F. Sattin[#]

Consorzio RFX, Euratom-ENEA Association,

Corso Stati Uniti 4, 35127 Padova, Italy

**Abstract**

The superstatistics approach recently introduced by Beck [C. Beck and E.G.D. Cohen, Physica **A 322**, 267 (2003)] is a formalism that aims to deal in a unifying way with a large variety of complex nonequilibrium systems, for which spatio-temporal fluctuations of one intensive variable ("the temperature" $1/\beta$) are supposed to exist. The intuitive explanation provided by Beck for superstatistics is based on the ansatz that the system under consideration, during its evolution, travels within its phase space which is partitioned into cells. Within each cell, the system is described by ordinary Maxwell-Boltzmann statistical mechanics, *i.e.*, its statistical distribution is the canonical one $e^{-\beta E}$, but $\beta$ varies from cell to cell, with its own probability density $f(\beta)$. In this work we first address that the explicit inclusion of the density of states in this description is essential for its correctness. The correction is not relevant for developments of the theory, but points to the fact that its correct starting point, as well its meaning, must be found at a more basic level: the pure probability product rule involving the intensive variable $\beta$ and its conjugate extensive one. The question therefore arises how to assign a meaning to these probabilities for each specific problem. We will see that it is easily answered through Bayesian analysis. This way, we are able to provide an interpretation for $f(\beta)$, that was not fully elucidated till now.

---

[#] e-mail: fabio.sattin@igi.cnr.it

## 1. Introduction

The superstatistics approach recently introduced by Beck [1] is a formalism that aims to deal in a unifying way with a large variety of complex nonequilibrium systems, for which spatio-temporal fluctuations of one (or possibly more) intensive variables ("the temperature" $1/\beta$) are supposed to exist. These fluctuations reverberate on the conjugate extensive variable ("the energy" $E$): its statistical distribution departs from the simplest canonical distribution. This formalism can accommodate, in principle, a multitude of empirically found anomalous statistical distributions: power-laws (indeed, they were the original motivation for developing this formalism), stretched exponentials, etc … . The literature about superstatistics is quickly increasing: the formalism, so far, has been used to interpret data from fluid turbulence [2-6], random matrix theory [7], astrophysics [8,9], just to mention some applications.

The intuitive explanation provided by Beck for superstatistics is based on the ansatz that the system under consideration, during its evolution, travels within its phase space which is partitioned into small cells. Within each cell, the system is described by ordinary Maxwell-Boltzmann statistical mechanics, *i.e.*, its statistical distribution is the canonical one $e^{-\beta E}$, but temperature-hence $\beta$-varies from cell to cell. Alternatively, one can think of $\beta$ as uniform throughout the whole phase space, but varying in time. The resulting, measurable statistics is therefore an average over the statistical distribution for $\beta$, $f(\beta)$:

$$P(E) = \int e^{-\beta E} f(\beta) d\beta \qquad (1)$$

This is the interpretation of superstatistics usually provided for didactical purposes [2]. For the sake of accuracy, Beck pointed it out that Eq. (1) is not properly normalized, and that a better way of writing it should be

$$P(E) = \int \frac{e^{-\beta E}}{Z(\beta)} f(\beta) d\beta \quad , \quad Z(\beta) = \int e^{-\beta E} dE \qquad (2)$$

so that $e^{-\beta E}/Z(\beta)$ is a proper probability density. The expressions (1) and (2) are referred to as respectively type-A and type-B superstatistics. It is clear, however, that this is just tantamount to a redefinition of $f(\beta)$ between (1) and (2). So, we will refer just to (1) in the following.

Eq. (1) has the merit of being simple and suggestive. Unfortunately, in this form, it is not correct. Let us see why: In a canonical ensemble whose phase space is (**q**,**p**), the probability for the system to be found within the small volume d**q** d**p** is

$$P(\mathbf{q},\mathbf{p}) \propto e^{-\beta H(\mathbf{q},\mathbf{p})} d\mathbf{q} d\mathbf{p} \qquad (3)$$

In order to compare (1) and (3) we need to convert the latter to a function of energy *E*: it is necessary a transformation of variables that introduces the density of states at energy *E*, $\omega(E)$: the Jacobian of the transformation ([10], ch. 7). $\omega(E)$ does not appear into (1) (nor in the partition function *Z* in Eq. 2, as it should be) and prevents this formula to be useful for dealing even with the simplest Gaussian case. Infact, the choice $E = \frac{1}{2} x^2$, $f(\beta) = C(\beta_0) \delta(\beta - \beta_0)$, with $\beta_0$ fixed constant and $\delta$ Dirac delta, does not lead to a Gaussian distribution because of an extra *x* factor coming from the Jacobian of the transformation $E \rightarrow x$.

Consistency may only reappear under the hypothesis that each element of the ensemble is evolving on a hypersurface of constant energy, $H = E$, hence $\omega(E)$ is constant and may be safely omitted from the integral. But this is not compatible with the canonical ensemble picture. The presence or absence of the density of states is therefore trivial mathematically in such equations as (1) or (2), but has a great conceptual importance.

For practical purposes, the situation is, however, not so much serious: Eq. (1) is actually never used when moving to actual computations, nor by Beck himself. The correct formulation of superstatistics is as follows: let us consider a scalar variable $\chi$ we are interested in. Superstatistics assumes there exists a conjugate variable $\beta_\chi$ such the sampling distribution for $\chi$ depends upon $\beta_\chi$ as well: $p(\chi) \equiv p(\chi, \beta_\chi)$. The joint probability is then factored into the conditional probability for $\chi$ and the marginal probability for $\beta_\chi$:

$$\begin{aligned} p(\chi,\beta_\chi) &= p(\chi|\beta_\chi) \times p(\beta_\chi) \\ \rightarrow p(\chi) &= \int p(\chi|\beta_\chi) \times p(\beta_\chi) d\beta_\chi \end{aligned} \qquad (4)$$

The second line in (4) yields the marginal probability for getting $\chi$ regardless of the value for $\beta_\chi$. The Gaussian case for $p(\chi|\beta_\chi)$ leads to

$$p(\chi) = \int \frac{1}{\sqrt{2\pi\beta_\chi^2}} \exp\left(-\frac{1}{2}\frac{\chi^2}{\beta_\chi^2}\right) \times p(\beta_\chi) d\beta_\chi \qquad (5)$$

and in this form superstatistics is ordinarily used. Beck himself introduced superstatistics through this form in his first work on the subject ([11], Eq. 15). In that paper he derived superstatistics from a dynamical realization of the underlying stochastic process, through a Langevin equation. It is well known that an ordinary

Langevin equation for a generalized "velocity" variable $u$ yields a stationary Gaussian distribution for $u$, which can formally be associated to an exponential for the "energy" $W$ through the relation $W = u^2/2$ [12,13]. However, the identification is only formal, since all formulas are still written in the "velocity" (phase) space, not in the energy space.

At this stage, any connection with thermodynamics has disappeared; Eq. (4) quantifies simply a logical relationship between probabilities. As such, the analytical form of both terms appearing in (4) is no longer constrained, nor for $p(\chi|\beta_\chi)$ itself: it not even needs to be an exponential. The relationship between $\chi$ and $\beta_\chi$ may be described by a variety of analytical expressions, of which the Gaussian case used in (5) is just a particular choice-although there are several reasons to endowe it with a special status. The precise form for $p(\chi|\beta_\chi)$ is usually not known beforehand, as instead implied in (1), but should be worked out case-by-case on the basis of the information at hand.

In studies based upon superstatistics a large interest is placed upon the functional form for $p(\beta_\chi)$, that is critical is determining the final $p(\chi)$: the original choice [1] was a chi-squared distribution, that leads to a power-law for $p(\chi)$. A log-normal function appears a plausible candidate in turbulence studies [5,14,15], and other distributions were also studied [2,16-18]. At this stage, it is difficult to choose $p(\beta_\chi)$ on the basis of some prior theory, and one is led mainly by the sought agreement *a posteriori* with experiment. Hence, no connection is drawn between $p(\chi|\beta_\chi)$ and $p(\beta_\chi)$ distributions: the former is postulated or deduced in advance of the problem, the latter guessed on the basis of the sought agreement with the experiment. But Eq. (4) requires $\chi$ and $\beta_\chi$ to be related for $p(\chi|\beta_\chi)$ not to become trivial: $p(\chi|\beta_\chi) \equiv p(\chi)$. In Eq. (5), e.g., $\beta_\chi$ is the variance for the distribution of $\chi$. In the past years, Lavenda and coworkers attacked fairly harshly Beck's works and pointed it out that a probabilistic description of statistical mechanics shows that fluctuations in energy and temperature are not independent [19-23] thermodynamic coniugate variables cannot be measured with infinite precision, errors in the measurement of one variable can be made vanishingly small only at the expense of total ignorance about the other. But finite errors in the estimate of a variable translate into a statistical distribution for its guessed value. Hence, not only statistical distribution for both variables arise naturally, but also they must be related to each other. (Incidentally, an embryonal version of this statement may be found in the paper [24]- see discussion around Eq. 13-written by this author without any prior knowledge of Lavenda's work).

At this point, a coherent picture is starting to emerge, enframed by the two facts that we are going to summarize: I) $p(\chi|\beta_\chi)$ is not an universal function as postulated into Eq. (1); it is a distribution that must be determined on the basis of the information that we have about the specific problem we are going to solve. II) $p(\beta_\chi)$, too, is not known in advance. Any prior knowledge we have about it must be modified in the light of experiment, that involves sampling on the coniugate variable. Notice that what we have just given here is nothing but a restatement of Bayes' formula (see, e.g., ref. [25], Eq. 1.3; or [26], Eq. 4.3).

From these two facts there follows a consideration: the emphasis is being naturally shifted from an "objective" picture (the system travels across different parts of its phase space) to a "subjective" one. While in (1) one has been emphasizing $f(\beta)$ to be a faithful description of a background truly acting stochastically, the lines above seem to suggest that a more productive way of looking at Eqns. (4,5) is to consider them not descriptive of a microscopical stochastic dynamics. Rather, they are representative of our *state of knowledge* about the system. The spreading of the statistical distributions over finite supports is merely a measure of how much our ignorance about fine details of the system (e.g., initial conditions, or boundary conditions) precludes us to univocally assigning numerical values to the quantities we want to measure.

What we have been describing in the lines above is just the Bayesian approach to inference as explained in his monograph by Jaynes [26], that will be our reference here (Bayesian method itself is, of course, much older, and is traced back essentially to Laplace). Bayesian probability theory is a method for drawing inferences using a well-defined set of logical axioms and rules (Cox's rules). Bayesian theory does not assign probabilities the status of physical entities: there does not exist anything like a random variable in the sense of a quantity that, objectively, picks randomly its value from an ensemble of possible choices. Rather, a probability represents our degree of belief about the value of the quantity under consideration, determined by our state of knowledge. This apparently may lead to purely subjective inferences, but once the state of knowledge has been mathematically quantified, it is perfectly legitimate to make quantitative comparisons between different predictions. Internal rules to the theory assure that, when logically equivalent states of knowledge are given, they lead to the same final results. A different, opposite, approach is the "ortodox" one, where probabilities are thought to represent true physical properties of the system studied, and can be quantified in the limit-of-frequency sense. Intermediate positions are also possible. A monograph about the foundational issues of the concept of probability in statistical physics is Guttmann [27].

In this paper we will carry on an interpretation of superstatistics using Bayesian theory. We will show how the probabilities entering Eq. (4) are given a precise formulation once the form of the information we have about the system is quantitatively established. We will show how, this way, one is able to recover explicitly Eq. (4) for $p(\beta_\chi)$ Gamma and a lognormal distributions, i.e., the two most important practical cases.

## 2. Recovering statistical distributions

Our task is recovering a probability distribution for $\chi$ on the basis of the measurable interaction of the system with its environment. By "measurable interaction" we mean that we expect the fine details of the interaction to escape or to be of no relevance to our measurements, and only coarse quantities may be retained, usually in the form of moments of the distribution function. Since in Eqns. (1,5) only quantities proportional to moments up to the second ($\beta, \beta_\chi$) do appear, we will stop to this order. We label the second moment with $\beta_\chi^2$. Furthermore, through a suitable shift of the reference frame, e.g., suitably choosing the origin of axes, we can always make the first moment to vanish. We stress that, up to this stage, $\beta_\chi$ is not a *measured* quantity. Our state of knowledge about the system tells us that it must be taken into account when deriving the probability distribution for $\chi$, but its precise value is of no concern to us. It is a nuisance parameter that will eventually be eliminated through marginalization (Eq. 4). Our task is that of giving explicit expressions for the two densities appearing in (4). Getting an expression for $p(\chi|\beta_\chi)$ is straightforward: by construction, the only information we have about it are its first two moments. The problem of transferring information uniquely into a probability is a central one in Bayesian theory, and is solved through a variety of approaches. Among them, one of the best known and most used is the Maximum Entropy Principle ([26], ch. 11): it tells us that the distribution to be preferred among those satisfying this constraint is given by solving the variational equation

$$\delta\left(\int p \ln\left(\frac{p}{\mu}\right) d\chi + \frac{1}{2\beta_\chi^2}\int p\chi^2 d\chi\right) = 0 \tag{6}$$

with $\int p d\chi = 1$ and $\mu(\chi)$ is a 'measure' function, needed to make $p/\mu$ invariant under a change of variables. It is easy to show that $\mu(\chi)$ is the prior distribution for $\chi$, in the lack of any other information. This can be spotted by solving (6) without even the term proportional to $1/\beta_\chi^2$: the result is $p \equiv \mu$. But in absence of any information we have no reason to prefer a value of $\chi$ over any other, hence $\mu$ must be a constant [the problem of infinite limits of integration in (6) is bypassed by noticing that no finite physical system may spread over infinite intervals. Hence, the integral must have some-possibly very large-cutoff]. The final result from (6) is

$$p(\chi|\beta_\chi) = \sqrt{\frac{1}{2\pi\beta_\chi^2}} \exp\left(-\frac{\chi^2}{2\beta_\chi^2}\right) \tag{7}$$

On the basis of the Maximum Entropy Principle, Eq. (7) is the most general (i.e., less informative) statistical distribution compatible with the constraint over its second moment. Notice that the functional to be maximized subject to constraints is the Shannon entropy, $-\int p \ln p \, dx$. Due to its analytical structure, any constraint that can be written in the form of an average over $p$ automatically leads to a probability density that is in the form $\exp(f(x))$. Hence, the leading role played by Gaussian distributions is easily explained.

Let us now investigate the second term, $p(\beta_\chi)$. $\beta_\chi$ was introduced on the basis of some measurement done, i.e., some data collected. Hence, the Bayes formula is needed ([25], Eq. 1.3; [26], Eq. 4.3):

$$p(\beta_\chi|D) = \frac{p(D|\beta_\chi) \times p(\beta_\chi|I)}{\int p(D|\beta_\chi) \times p(\beta_\chi|I) d\beta_\chi} \tag{8}$$

The meaning of (8) is: the probability for $\beta_\chi$ conditional to some information $D$ is given by the probability of measuring $D$ for fixed $\beta_\chi$, $p(D|\beta_\chi)$, multiplied by the prior (i.e., in absence of the information $D$) probability we assign to $\beta_\chi$, $p(\beta_\chi|I)$. All must be suitably normalized to yield a proper probability.

The physics enters through $D$. In his first paper on the subject [11], Beck makes the hypothesis that the "kinetic energy" $\frac{1}{2}\chi^2$ may be written as the sum of three independent Kolmogorov velocities: $1/2\chi^2 = 1/2\sum_{i=1}^{3}\xi_i^2$. We can generalize this ansatz to an arbitrary number *m* of components:

$$\chi^2 = \sum_{i=1}^{m}\xi_i^2 \qquad (9)$$

In order to estimate $\beta_\chi$ we need to have performed some measure of the energy of the system or, from (9), of the *m* velocity components. Hence $p(D|\beta_\chi)$ stands for the joint probability density:

$$p(D|\beta_\chi) = \left(\frac{1}{2\pi\beta_\chi^2}\right)^{m/2} \prod_{i=1}^{m} \exp\left(-\frac{\xi_i^2}{2\beta_\chi^2}\right) \qquad (10)$$

The factorization comes straightforwardly by the hypothesis of the $\xi_i$ to be independent components, and the Gaussian statistics arises for the same reasons the led to (7).

In (8) only probability densities do enter, but we can convert $p(D|\beta_\chi)$ to a true probability by multiplying both numerator and denominator by the volume element $\prod_i d\xi_i$. This probability must be invariant under the change of variables: $(\xi_1, \xi_2, \ldots)$ → $(X, \theta_2, \ldots)$, where $\theta_i$ ($i = 2, \ldots, m$) are angle coordinates and $\sum_{i=1}^{m}\xi_i^2 \to X$. We arrive to

$$p(D|\beta_\chi) \propto \left(\frac{1}{2\pi\beta_\chi^2}\right)^{m/2} (X)^{m/2-1} \exp\left(-\frac{X}{2\beta_\chi^2}\right) \qquad (11)$$

The factor $X^{m/2-1}$ comes from the jacobian of the transformation. Other constant terms are not relevant to our purposes, since they are cancelled between numerator and denominator.

There remains, finally, the prior probability $p(\beta_\chi/I)$: the Jeffreys rule ([26], ch. 12.4) states that, in absence of whatsoever information, the prior probability for a positive definite quantity *h* must be taken as $p(h/I) \propto 1/h$: it is a necessary consequence of the invariance of our knowledge about *h* after transformation of the scales of the problem (For the unfamiliar reader, a short proof of this result is given in the Appendix).

Inserting everything into (8) yields

$$p(\beta_\chi | D) = \frac{X^{m/2}}{2^{m/2-1}\Gamma(m/2)} \frac{1}{\beta_\chi^{m+1}} \exp\left(-\frac{X}{2\beta_\chi^2}\right) \qquad (12)$$

The quantity $X$ is arbitrary but, for a thermodynamical system, with overwhelming probability, any measure will yield approximately the same, most likely value: $X = \bar{X}$ ([10], par. 7.2). In order to get a closer comparison with Beck's results, let us perform the change of variables: $\bar{X}/2 \to 1/b_0, 1/\beta_\chi^2 \to b$. After replacing (7, 12) into (4) we get, finally

$$p(\chi) = \int \frac{1}{\sqrt{2\pi}\Gamma(m/2)} \frac{1}{b_0^{m/2}} b^{\frac{m-1}{2}} \exp\left(-\frac{b}{b_0}\right) \exp\left(-b\frac{\chi^2}{2}\right) db \qquad (13)$$

which is the convolution of a Gaussian with a Gamma distribution. This yields, as known, a power law for $p(\chi)$ [11]. Notice that, with respect to original Beck's guess, there is a supplementary "1/2" in the exponent for $b$: $m/2 - 1/2$ rather than $m/2 - 1$. This comes from our choice of prior probability $p(\beta_\chi/I)$. Beck's ansatz is consistent, instead, with $p(\beta_\chi/I) = $ constant. (although Beck himself, later, recovered the $m/2 - 1/2$ using B-superstatistics rather than A-superstatistics [2]).

The other relevant, often used distribution for $p(\beta_\chi)$ is the log-normal one [2]. This may be recovered from the ansatz that the variable $\chi^2$ be written as a *product* of $m$ independent $\xi$ variables:

$$\chi^2 = \left(\prod_i \xi_i\right)^{1/m} \to \ln\left(\frac{\chi^2}{c\beta_\chi^2}\right) = \frac{1}{m}\sum_i \ln\left(\frac{\xi_i}{c\beta_\chi^2}\right) \qquad (14)$$

The parameter $c$ is chosen on the basis of the constraint that $\ln(\chi^2/c\beta_\chi^2)$ be zero-mean: $c = \exp(-\gamma_E)/2$, with $\gamma_E \approx 0.577\ldots$ . Hence $\ln(\xi_i/c\beta_\chi^2)$ are zero-mean, too. Again, the only information we postulate to have is about their second moment, and this leads (again on the basis of Maximum Entropy Principle) to

$$p(\xi_i | \beta_\chi) = \frac{1}{\sqrt{2\pi s^2}} \exp\left[-\frac{1}{2}\frac{(\ln(\xi_i/c\beta_\chi^2))^2}{s^2}\right] \qquad (15)$$

The variance $s$ is assigned:

$$s^2 = \left\langle\left(\ln\left(\frac{\xi_i}{c\beta_\chi^2}\right)\right)^2\right\rangle$$

$$\to s^2 = \frac{1}{m}\sum_i \left\langle\left(\ln\left(\frac{\xi_i}{c\beta_\chi^2}\right)\right)^2\right\rangle = \left\langle\left(\ln\left(\frac{\chi^2}{c\beta_\chi^2}\right)\right)^2\right\rangle \qquad (16)$$

$$= \gamma_E^2 + \frac{\pi^2}{2} + (\ln(2))^2 + \gamma_E \ln(4) + \ln(c)[2\gamma_E + \ln(4c)] = \frac{\pi^2}{2}$$

where the average is done over the gaussian probability density for $\chi$ (Eq. 7).

On the basis of the same reasoning done before, for a thermodynamical systems, all the $\xi_i$ are almost constant and identical: $\xi_i/c \equiv \beta_0^2$, and Eq. (8) becomes

$$p(\beta_\chi | D) = \left(\frac{m}{2\pi s^2}\right)^{1/2} \exp\left[-\frac{m}{2s^2}\left(\ln\left(\frac{\beta_x}{\beta_0}\right)^2\right)^2\right]\frac{1}{\beta_\chi} \qquad (17)$$

Inserting this expression into (4) together with (7), and with the replacements $1/\beta_\chi^2 \to b, 1/\beta_0^2 \to b_0$, leads to

$$p(\chi) = \frac{\sqrt{m}}{2\pi s}\int db\,\frac{1}{b^{1/2}}\exp\left(-\frac{b}{2}\chi^2\right)\exp\left[-\frac{m}{2s^2}\left(\ln\left(\frac{b}{b_0}\right)\right)^2\right] \qquad (18)$$

concluding our derivation.

It is interesting to notice that, independently of the considerations put forth in this work, the Bayesian method outlined in this paragraph has already actually put into practice within superstatistical theory: see the derivation of statistical distributions from Maximum Entropy Principle done by Reynolds [5] and by the present author [17].

## 3. Concluding remarks

Beck's superstatistics cannot be considered completely new: the mechanism of superposition of probability densities to obtain "anomalous" (i.e., non-Gaussian) distributions is well known in statistics as well in physics (see, e.g., [28], par. 14.4): indeed, the first suggestion to a formula like (1) came from the paper [29]. Lavenda and Dunning-Davies [30] claim that even the results of this latter paper should be traced back to earlier work. It is possible that these results have been actually discovered and used several times independently by several researchers in the past (see, e.g., [31]). We think, instead, that Beck's attempt of finding a common intuitive foundation for a large variety of apparently disconnected experimental results, is valid and promising, and deserves to be further developed. We hope that the work presented in this paper might be a contribution to superstatistics theory.

This work has two goals: first, the original interpretation (1) is found to need amendements. Therefore, it has been pointed out that the more correct starting point is the product rule, Eq. (4) (Notice that we are not claiming originality, here: Eq. (4) was already suggested by Beck). An interpretation of the probabilities there appearing within the Bayesian picture allows us to recover almost effortless the main results of superstatistics theory. We consider putting the emphasis on the Bayesian approach for all of the probabilities appearing into Eq. (4) a distinctive feature of this work. It allowed to address a fundamental issue otherwise not extensively investigated till now, namely the mechanisms through which statistical distributions for the intensive variable $\beta$ arise from within the systems studied.

**Acknowledgments**

Part of this paper was stimulated by discussions with L. Salasnich. Professors Beck and Lavenda made useful suggestions, although not necessarily sharing the contents of the paper. This work was supported by the Euratom Communities under the contract of Association between EURATOM/ENEA. The views and opinions expressed herein do not necessarily reflect those of the European Commission.

**Appendix: The Jeffrey's prior**

A concise derivation of Jeffrey's prior may be found, e.g., in Sivia ([25], p. 112). For the reader's sake, we provide here essentially a copy of that result.

We wish to assign a functional form to the PDF $p(h/I)$, where $h$ represents the quantity of interest and $I$ stands for any other information related to the problem. We have not any information about $h$ but for the fact that it must be positive: $h > 0$.

Gross ignorance about $h$ means complete lack of knowledge about scales involved in the problem, or, equivalently, invariance of the probability: in fact, a change of $p$ as a consequence of a change of scales (which could simply be changing the units with which we measure $h$) means automatically that we have some information about it, explicitly denied from the outset. A change of scales corresponds to a stretching or shrinking of $h$. Hence, the probability must fulfil:

$$p(h|I)dh = p(\alpha h|I)d(\alpha h) \quad , \quad \alpha > 0$$

and this can only be satisfied if

$$p(h|I) \propto 1/h$$

which concludes the proof.

**References**

[1] C. Beck and E.G.D. Cohen, Physica **A 322**, 267 (2003).

[2] C. Beck, in Proceedings of of the 31st Workshop of the International School of Solid State Physics, Complexity, Metastability And Nonextensivity, Erice (Sicily), 20-26 July 2004 (Eds. C. Beck, G. Benedek, A. Rapisarda and C. Tsallis, World Scientific, 2005) [cond-mat/0502306].

[3] C. Beck, Physica **A 342**, 139 (2004)

[4] C. Beck, Physica **D 193**, 195 (2004)

[5] A.M. Reynolds, Phys. Rev. Lett. **91**, 084503 (2003)

[6] S. Rizzo, A. Rapisarda, in Proceedings of of the 31st Workshop of the International School of Solid State Physics, Complexity, Metastability And Nonextensivity, Erice (Sicily), 20-26 July 2004 (Eds. C. Beck, G. Benedek, A. Rapisarda and C. Tsallis, World Scientific, 2005) [cond-mat/0502305]

[7] A.Y. Abul-Magd, Physica **A 361**, 41 (2006)

[8] M. Baiesi, M. Paczuski, A.L. Stella, cond-mat/0411342 (2004)

[9] P.-H. Chavanis, Physica **A 359**, 177 (2006)

[10] K. Huang, *Statistical Mechanics*, 2$^{nd}$ Ed. (John Wiley & Sons, 1987).

[11] C. Beck, Phys. Rev. Lett. **87**, 180601 (2001).

[12] W.T. Coffey, Yu P. Kalmykov, and J.T. Waldron, *The Langevin equation* (World Scientific, 1996), ch. 1.

[13] S. Chandrasekhar, Rev. Mod. Phys. **15**, 1 (1943). Reprinted in *Selected Papers*, vol. 3 (The University of Chicago Press, 1989).

[14] C. Jung, H.L. Swinney, Phys. Rev. E **72**, 026304 (2005)

[15] A.K. Aringazin and M.I. Mazhitov, Phys. Lett. **A 313**, 284 (2003).

[16] F. Sattin and L. Salasnich, Phys. Rev. E **65**, 035106(R) (2002).

[17] F. Sattin, Phys. Rev. E **68**, 032102 (2003).

[18] F. Sattin, Physica **A 338**, 437 (2004).

[19] B.H. Lavenda and J. Dunning-Davies, cond-mat/0311271 (2003).

[20] B.H. Lavenda and J. Dunning-Davies, cond-mat/0312301 (2003).

[21] B.H. Lavenda, cond-mat/0408485 (2004).

[22] B.H. Lavenda, *Statistical Physics* (John Wiley & Sons, 1991).

[23] B.H. Lavenda, cond-mat/0401024 (2004).

[24] F. Sattin, J. Phys. A: Math. Gen. **36**, 1583 (2003).

[25] D.S. Sivia, *Data Analysis* (Clarendon Press, 1996).

[26] E.T. Jaynes, *Probability Theory* (Cambridge University Press, 2003).

[27] Y.M. Guttmann, *The Concept of Probability in Statistical Physics* (Cambridge University Press, 1999).

[28] D. Sornette, *Critical Phenomena in Natural Sciences* (Springer, 2000).

[29] G. Wilk and Z. Wlodarczyk, Phys. Rev. Lett. **85**, 2770 (2000).

[30] J. Dunning-Davies, physics/0502153 (2005).

[31] G.M. Batanov *et al*, Plasma Phys. Rep. **28**, 111 (2002).